\documentclass[aps,preprint,amsmath,amssymb,prb,superscriptaddress,floatfix,dvipdf]{revtex4-1}
\usepackage[dvipdf]{graphicx}
\usepackage{dcolumn}
\usepackage{bm}
\usepackage{braket}
\usepackage{color}

\allowdisplaybreaks[4] 

\begin{document}


\title{Anisotropic superconducting gaps in YNi$_2$B$_2$C:
  A first-principles investigation}



\author{Mitsuaki Kawamura}
\email{mkawamura@issp.u-tokyo.ac.jp}
\affiliation{Institute for Solid State Physics, 
  The University of Tokyo, Kashiwa 277-8581, Japan}

\author{Ryosuke Akashi}
\affiliation{Department of Physics, The University of Tokyo, Tokyo 113-0033, Japan}

\author{Shinji Tsuneyuki$^{1,}$}
\affiliation{Department of Physics, The University of Tokyo, Tokyo 113-0033, Japan}



\date{\today}

\begin{abstract}
  We calculate superconducting gaps and quasiparticle density of states
  of YNi$_2$B$_2$C 
  in the framework of the density functional theory for superconductors
  to investigate the origin of a highly anisotropic superconducting gaps 
  in this material.
  Calculated phonon frequencies, the quasiparticle density of states, 
  and the transition temperature show good agreement with experimental results.
  From our calculation of superconducting gaps and orbital character analysis,
  we establish that the orbital character variation of the Fermi surface is the 
  key factor of the anisotropic gap.
  Since the electronic states that consist of mainly 
  Ni $3 d$ orbitals couple weakly with phonons,
  the superconducting gap function is suppressed for the corresponding states, 
  which results in the anisotropy observed in the experiments.
  These results are hints to increase the transition temperature of materials 
  in the borocarbide family.
\end{abstract}


\maketitle
%
%
\section{Introduction}
Superconductors exhibiting anisotropic gap has attracted continuous attention for its possible exotic superconducting mechanisms. 
Unconventional mechanisms,
which is completely different from the phonon mechanism established by
the Bardeen-Cooper-Schrieffer(BCS)\cite{PhysRev.108.1175} and Eliashberg\cite{SovPhysJETP.11.696} theories,
have been extensively discussed for the nodal $d$- and $p$-wave gaps in cuprate, 
iron-based, and heavy-fermion superconductors
\cite{ISI:000282019500009,RevModPhys.83.1589,RevModPhys.81.1551}
, etc. 
In view of this history, YNi$_2$B$_2$C\cite{Mazumdar1993413,ISI:A1994MR49400048}
is recently getting a surge of interest since its superconducting gap is significantly anisotropic. 
Polynomial temperature dependence of its specific-heat coefficient has been observed ($C_{p}\propto T^3$), 
suggesting nodal structure of the gap. 
The strong anisotropy of the gap has also been indicated 
in the magnetic-field dependence of $C_{p}$~\cite{doi:10.1143/JPSJ.68.1078,Nohara20002177,PhysRevLett.86.1327}, 
broad peak in the tunneling conductance spectrum~\cite{PhysRevB.67.014526}, 
in-plane anisotropy in the ultrasonic attenuation~\cite{PhysRevLett.92.147002} and Doppler shift measurements~\cite{PhysRevLett.89.137006}, 
and large anisotropic gap ratio ($\Delta_{\rm max}/\Delta_{\rm min} = 2.1$),
namely, the ratio of the maximum to the minimum of the gap in the reciprocal space, observed 
with Angle-resolved photoemission spectroscopy measurement~\cite{PhysRevB.81.180509}. 
Strong antiferromagnetic spin fluctuation has been revealed from pulsed NMR studies~\cite{PhysRevB.51.3985}, 
which suggests that electronic correlation has a role.

Although significant magnetic characteristics are generally observed
in the rare-earth nickel borocarbide family,
the yttirium systems seems exceptional.
Among $Ln$Ni$_2$B$_2$C ($Ln$=lanthanide),
the Pr\cite{489885}, Nd\cite{NAGARAJAN1995571}, Sm\cite{0295-5075-29-8-009},
Gd\cite{Canfield1995337} and Tb\cite{PhysRevB.53.8499} systems exhibit magnetic order,
whereas in the Dy\cite{PhysRevB.52.R3844}, Ho\cite{PhysRevB.50.9668},
Er\cite{PhysRevB.51.678} and Tm\cite{PhysRevB.52.3676} systems
both magnetic orders and superconducting transition have been observed.
Previous first-principles investigations\cite{ZENG199623, 0953-8984-18-26-016} revealed that
the magnetic orders in those materials are caused by
the Ruderman-Kittel-Kasuya-Yosida interaction\cite{PhysRev.96.99,Kasuya01071956,PhysRev.106.893}
between localized spin from 4f electrons.
In the Y system, on the other hand, 
the yttrium sites, whose valence states being less localized 4d and 5s orbitals,
does not show magnetic order.
This implies that the effective description of electronic states in the Y systems
should be different from the other rare-earth systems.

As a matter of fact,
a relevance of the conventional phonon-mediated superconducting mechanism has also been experimentally indicated.  
First, the isotope effect of boron atom have been observed~\cite{Lawrie1995159,Cheon199935} in this material. 
Moreover, softening of the transverse acoustic (TA) phonon mode occurs, 
which is likely due to strong electron-phonon coupling\cite{PhysRevB.89.104503}. 
The apparent coexistence of the strong electron-phonon coupling and the gap anisotropy invites us to a fundamental question: 
Can the conventional superconducting mechanism realize such anisotropic gap? 
Although the conventional mechanism is generally regarded to induce almost isotropic gap\cite{ehrenreich1983solid}, 
there is no theory to prohibit the opposite. In a few multiband systems such as MgB$_2$~\cite{ISI:000167194300040, ISI:000177428000033}, 
the gap has different values for different bands, 
which can be explained with orbital dependence of the electron-phonon coupling. 
Even the nodal gap can theoretically emerge 
if we assume extreme $k$-point dependence of the electron-phonon coupling\cite{PhysRevLett.77.723}.

In this study, we investigate
the possibility of the anisotropic superconductivity due to the conventional phonon mechanism
in YNi$_2$B$_2$C in a fully {\it ab initio} manner.
Recent progress in {\it ab initio} theories for superconductors 
such as density functional theory for superconductors (SCDFT) \cite{PhysRevLett.60.2430, PhysRevB.72.024545} 
and anisotropic Migdal-Eliashberg theory\cite{PhysRevB.87.024505}, has enabled us to work on this issue.
The standard method to calculate gaps of the superconducting phase 
induced by the phonon-mediated mechanism is to solve the Eliashberg equation.
However, it is difficult to solve it in fully non-empirical manner;
because the interaction and the gap function depend both on the Kohn-Sham state and the frequency, 
formidable computational cost is required for solving this equation.
On the other hand, in the density functional theory for superconductors, 
static anomalous density serves as a fundamental quantity, 
with which efficient numerical schemes can be implemented with reduced computational cost.
In the recent SCDFT study, 
$T_{\rm c}$ as well as the tunneling gap have been reproduced from first principles
\cite{PhysRevB.75.020511}.
We apply this method to YNi$_2$B$_2$C system.

In Sec. II, we introduce the density functional theory for superconductors, which bases our first-principles calculations in this study. 
%
In Sec. III, we present the numerical method to calculate superconducting properties such as the superconducting gap. 
We here append a specific scheme to treat the $k$-dependent anisotropy precisely.
%
In Sec. IV, 
we show the resulting electronic and phononic structure of YNi$_2$B$_2$C;
the band structure, Fermi surfaces, superconducting gap function, phonon dispersion, superconducting transition temperature, 
and the quasiparticle density of states. 
We compare them with the corresponding experimental results.
In Sec. V, we discuss the possible origin of the anisotropic superconducting gaps in YNi$_2$B$_2$C 
within the conventional phonon-mechanism scenario.
%
The summary and future prospects are given in Sec. VI.

%
%
\section{Theory}


In this section, we present the formalism of
density functional theory for superconductors (SCDFT)
\cite{PhysRevLett.60.2430};
we can calculate superconducting properties such as the transition temperature
from first principles by using this method.

In the current formalism of SCDFT
with the decoupling approximation\cite{PhysRevB.72.024545, PhysRevB.72.024546}, 
we obtain the superconducting singlet order parameter as follows:
\begin{align}
  \chi(r, r') &= \frac{1}{2}\frac{\Delta_{n k}}{\varepsilon_{n k}}
  \tanh \left( \frac{\beta \varepsilon_{n k}}{2} \right)
  \sum_{n k} \varphi_{n k}(r)\varphi_{n k}^*(r'),
  \label{fml_scorder}
\end{align}
where $\varepsilon_{n k} = \sqrt{\xi_{n k}^2 + \Delta_{n k}^2 }$,
$\varphi_{n k}(r)$ is the normal-state Kohn-Sham orbital
having $(n k)$ as the band index and the wave number,
$\xi_{n k}$ is the normal-state Kohn-Sham eigenenergy,
and $\beta$ is the inversed temperature.
This order parameter is determined after we compute the electronic and lattice-dynamical properties
by using the density functional calculations in the normal state.
The Kohn-Sham gap function $\Delta_{n k}$ is calculated 
from the following gap equation:
\begin{align}
  &\Delta_{n k} = -\frac{1}{2} \sum_{n' k'} 
  K[V_{n' k' n k}](\{g_{n' k' n k}\}, \{\omega_{k'-k}\}, \xi_{n k}, \xi_{n' k'}) 
  \tanh \left ( \frac{\beta \varepsilon_{n' k'}}{2} \right ) 
  \frac{\Delta_{n' k'}}{\varepsilon_{n' k'}},
  \label{fml_ksgapeq}
  \\
  K[V_{n' k' n k}]&(\{g_{n' k' n k}\}, \{\omega_{k'-k}\}, \xi_{n k}, \xi_{n' k'}) 
  \equiv
  \frac{K^{\rm ep}(\{g_{n' k' n k}\}, \{\omega_{k'-k}\}, \xi_{n k}, \xi_{n' k'}) 
    + K^{\rm ee}[V_{n' k' n k}](\xi_{n k}, \xi_{n' k'}) }
       {1 + Z(\{g_{n' k' n k}\}, \{\omega_{k'-k}\}, \xi_{n k})},
\end{align}
where $\{\omega_{q}\} \equiv \omega_{q 1}, \cdots, \omega_{q N_{\rm branch}}$ 
are frequencies of phonons having $q$ as the wave number, 
$\{g_{n' k' n k} \} \equiv g_{n' k' n k}^{1}, \cdots, g_{n' k' n k}^{N_{\rm branch}}$
are the vertices between a phonon and 
Kohn-Sham orbitals $(\varphi_{n' k'}, \varphi_{n k})$,
$N_{\rm branch}$ is the number of branches of phonons.
Forms of the electron-phonon kernel $K^{\rm ep}$
and the renormalization $Z$ are identical to that of previous studies 
\cite{PhysRevLett.96.047003, PhysRevLett.111.057006, PhysRevB.75.020511}.
%
%
%
$K^{\rm ee}$ is the electron-electron kernel as follows:
\begin{align}
  K^{\rm ee}[V_{n' k' n k}](\xi_{n k}, \xi_{n' k'}) &= 
  \frac{2}{\pi} \int_0^{\infty} d \omega \frac{|\xi_{n k}|+|\xi_{n' k'}|}
       {(|\xi_{n k}|+|\xi_{n' k'}|)^2 + \omega^2} V_{n' k' n k}(\omega),
  \label{fml_kelkernel}
  \\
  V_{n' k' n k}(\omega) &\equiv \iint d^3r d^3r' \varphi^{*}_{n k}(r) \varphi^{*}_{n' k'}(r')
  V_{\rm scr}(r, r', \omega) \varphi_{n k}(r') \varphi_{n' k'}(r),
\end{align}
where $V_{\rm scr}(r,r',\omega)$ 
is the screened Coulomb interaction; 
we calculate it including the dynamical screening effect \cite{PhysRevLett.111.057006}.
Neglecting the temperature dependence of $K^{\rm ee}$ considered in the previous study
[Eq. (2) in the Ref. \onlinecite{PhysRevLett.111.057006}], 
we obtain Eq. (\ref{fml_kelkernel}).
The numerical treatment of the integration in Eq. (\ref{fml_kelkernel}) is
appended in Appendix \ref{app_omegaint}.

While $V_{\rm scr}(r,r',\omega)$ is calculated by using the random phase approximation (RPA) 
in the previous study\cite{PhysRevLett.111.057006},
we calculate it by using the adiabatic local density approximation\cite{PhysRevA.21.1561} (ALDA) in this work
as follows:
\begin{align}
  V_{\rm scr}(r,r',\omega) = \frac{1}{|r-r'|} 
  + \iint d^3 r_1 d^3 r_2
  \left( \frac{1}{|r - r_1|} + \frac{\delta^2 E_{\rm XC}}{\delta \rho(r) \delta \rho(r_1)} \right)
  \Pi(r_1, r_2, \omega) \frac{1}{|r_2 - r'|}
  \label{fml_screenedcoulomb}
\end{align}
where $\Pi(r, r', \omega)$ is the polarization function given by the solution of the following equation
\begin{align}
  \Pi(r, r', \omega) = \Pi_0(r, r', \omega) + \iint d^3 r_1 d^3 r_2
  \Pi_0(r, r_1, \omega) 
  \left( \frac{1}{|r_1-r_2|} + \frac{\delta^2 E_{\rm XC}}{\delta \rho(r_1) \delta \rho(r_2)} \right)
  \Pi(r_2, r', \omega),
\end{align}
and $\Pi_0(r, r', i \omega)$ is the independent particle polarizability
\begin{align}
  \Pi_0(r, r', \omega) = \sum_{n k n' k'} 
  \frac{\theta(-\xi_{n k}) - \theta(\xi_{n' k'})}{\xi_{n' k'} - \xi_{n k}+i \omega} 
  \varphi^*_{n k}(r) \varphi^*_{n' k'}(r') \varphi_{n k}(r') \varphi_{n' k'}(r),
  \label{fml_pi0}
\end{align}
where $\theta(\xi)$ is the step function. 

When we compare the calculated $\Delta_{nk}$ with the experiments, 
it must be noticed that $\Delta_{nk}$ is not theoretically guaranteed to correspond to the experimental gap; 
while the former gives the gap in the Kohn-Sham Bogoliubov-de Gennes energy dispersion,
the latter is defined with the poles of the electronic Green's function. 
Nevertheless, we discuss the gap anisotropy with the calculated $\Delta_{nk}$ 
on the basis of an assumption that it describes the experimental gap on the semiquantitative level. 
This is justified for the following reasons: (i) There is a suggestive relation between the SCDFT gap and those 
from the many body perturbation theory\cite{JLowTempPhys.10.79, doi:10.1143/JPSJ.45.786}.
From dressed anomalous Green's function $F^{R}_{n k}(\omega)$ in the Nambu-Gor'kov formalism \cite{PhysRev.117.648, ISI:A1958WX85400031},
let us redefine $\Delta_{nk}$ as follows:  
\begin{align}
  \Delta_{n k} \equiv 2 |\xi_{n k}| \int_0^\infty \frac{d \omega}{2 \pi} {\rm Im} F^{R}_{n k}(\omega).
\end{align}
Here, $\xi_{n k}$ is the Kohn-Sham energy of the normal state.
Remarkably, this redefined $\Delta_{n k}$ gap satisfies the equation similar to the SCDFT gap equation \cite{doi:10.1143/JPSJ.45.786}. 
(ii) The ``gap" $2\Delta_{nk}$ indeed agrees with the gap derived from the tunnel conductance for some materials~\cite{}. 
If one wants to improve the precision of the analysis, 
dressed anomalous Green'a functions must be calculated from first principles based 
on the Eliashberg theory \cite{SovPhysJETP.11.696, schrieffer1983theory}
or the $G_0 W_0$ theory for superconductors \cite{doi:10.1143/JPSJ.45.786, doi:10.1143/JPSJ.49.1267}, 
but it requires us to treat all the variables ($k$ points, band indices, 
and Matsubara frequencies) explicitly. Since the numerical cost for such a calculation is unrealistically large, 
we do not address this improvement in this study.

%

%
%
\section{Numerical method for the gap equation}

In this section, we explain the numerical method to compute the gap function [Eq.~(\ref{fml_ksgapeq})], the
independent particle polarizability [Eq.~(\ref{fml_pi0})], and the quasiparticle density of states in a superconducting state. 
All these calculations concern the $k$-point integrations where the integrands have large values only in the vicinity of the Fermi level. 
We developed a method based on the tetrahedron interpolation for this difficulty.

\subsection{Difficulty in the calculation of the gap equation}

\begin{figure}[bt]
  \includegraphics[width=8cm]{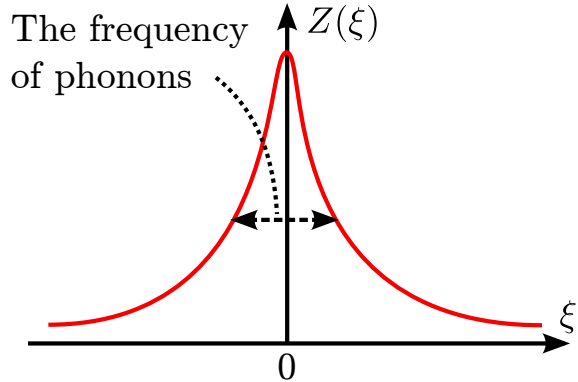}
  \caption{(Color online) Schematic illustration of the energy dependence of $Z$.
    This function has a strong energy dependence when $|\xi_{n k}|$ are equal to or lower 
    than the phonon frequency.
    \label{fig_eng_depend}}
\end{figure}

The renormalization factor $Z$
and the electron-phonon XC kernel $K^{\rm ep}$
vary rapidly in the vicinity of Fermi surfaces.
The origin of this rapid variation is strong energy ($\xi_{n k}$, $\xi_{n' k'}$) dependence 
of $Z$ and $K^{\rm ep}$ in the vicinity of Fermi surfaces
[$|\xi_{n k}|$ and $|\xi_{n' k'}|$ are equal to or lower 
than the phonon frequency. See Fig. \ref{fig_eng_depend}.];
%
in order to treat this sensitive energy dependence precisely,
we need an unrealistically large number of $k$ points
for solving the Kohn-Sham gap equation if we use the uniform grid.

In the previous works \cite{PhysRevB.72.024546}, 
randomly sampled $k$ points have been used to perform the $k$ integration
in the gap equation;
large number of $k$ points are adopted in the vicinity of Fermi surfaces.
However, this method has two drawbacks.
In the first place, it obviously yields a numerical error because of the random sampling.
In the second place, it has a difficulty in the calculation of the density of states
because we can not obtain exact weights of an integration
including the delta function;
for calculating such weights,
we have to use the tetrahedron method\cite{Jepson19711763}
on sufficiently dense regular $k$ grids (not on randomly sampled $k$ points).

\subsection{Deterministic solving via auxiliary gap function}
\label{sec:anxiliary-gap}
\begin{figure}[bt]
  \includegraphics[width=8cm]{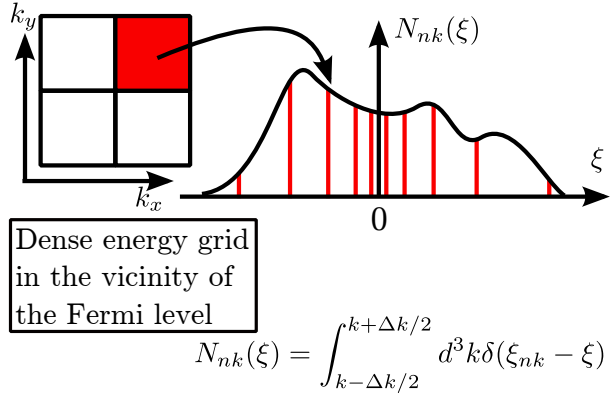}
  \caption{\label{fig_auxeng}
    (Color online) Schematic illustration of the auxiliary energy grid.
    We use smooth uniform $k$ grid and dense non-uniform energy grid to solve this 
    gap equation;
    the latter has much more points in the vicinity of $\xi = 0$.}
\end{figure}

To avoid this difficulty, we develop an alternative {\it deterministic} method 
that is free from the randomness and compatible with the tetrahedron method.
We decouple the $k$ dependence and energy dependence
with a help of the auxiliary energy grid.
Specifically, we define explicitly energy dependent auxiliary gap functions
\begin{align}
  \Delta_{n k} (\xi) &\equiv - \frac{1}{2} \sum_{n' k'} 
  K[V_{n' k' n k}](\{g_{n' k' n k}\}, \{\omega_{k'-k}\}, \xi, \xi_{n' k'}) 
  \tanh \left( \frac{\beta \varepsilon_{n' k'} }{2} \right) \frac{\Delta_{n' k'}}{\varepsilon_{n' k'}}.
  \label{fml_auxdelta}
\end{align}
This auxiliary gap function satisfy $\Delta_{n k}(\xi_{n k}) = \Delta_{n k}$.
Inserting $1 = \int d\xi' \delta(\xi' - \xi_{n' k'})$ into Eq. (\ref{fml_auxdelta}),
we obtain simultaneous equations for the auxiliary gap function as follows:
\begin{align}
  \Delta_{n k} (\xi) = - \frac{1}{2} \int d\xi' \sum_{n' k'} \delta(\xi' - \xi_{n' k'})
  K[V_{n' k' n k}](\{g_{n' k' n k}\}, \{\omega_{k'-k}\}, \xi, \xi') 
  \tanh \left( \frac{\beta \varepsilon_{n' k'}(\xi')}{2} \right) 
  \frac{\Delta_{n' k'}(\xi')}{\varepsilon_{n' k'}(\xi')},
  \label{fml_auxdelta2}
\end{align}
where $\varepsilon_{n k}(\xi) \equiv \sqrt{|\Delta_{n k}(\xi)|^2 + \xi^2}$.
We use a sparse uniform $k$ grid and non-uniform energy grid to solve this 
gap equation;
the latter has much more points in the vicinity of $\xi = 0$ (Fig. \ref{fig_auxeng}).

Practically, the energy dependence of $Z$ and $K^{\rm ep}$ 
becomes moderate when $\xi_{n' k'}$ is far from the Fermi level;
we therefore introduce the integration with respect to $\xi_{n'k'}$ 
only for bands crossing the Fermi level as follows:
\begin{align}
\hspace{-2.5cm}
  \Delta_{n k} (\xi) = &- \frac{1}{2} \int_{\xi_{\rm min}}^{\xi_{\rm max}}
  d\xi' \sum_{n'}^{\rm Fermi} \sum_{k'} \delta(\xi' - \xi_{n' k'})
  K[V_{n' k' n k}](\{g_{n' k' n k}\}, \{\omega_{k'-k}\}, \xi, \xi') 
  \tanh \left( \frac{\beta \varepsilon_{n' k'}(\xi')}{2} \right) 
  \frac{\Delta_{n' k'}(\xi')}{\varepsilon_{n' k'}(\xi')}
  \nonumber \\
  &- \frac{1}{2} \sum_{n'}^{\rm Other} \sum_{k'} 
  K[V_{n' k' n k}](\{g_{n' k' n k}\}, \{\omega_{k'-k}\}, \xi, \xi_{n' k'}) 
  \tanh \left( \frac{\beta \varepsilon_{n' k'}(\xi_{n' k'})}{2} \right) 
  \frac{\Delta_{n' k'}(\xi_{n' k'})}{\varepsilon_{n' k'}(\xi_{n' k'})},
  \label{fml_auxdelta3}
\end{align}
where $\xi_{\rm max}$ and $\xi_{\rm min}$ are the maximum of the normal-state Kohn-Sham energy 
of bands crossing the Fermi level and minimum of that.

%
For evaluating the $\xi$ integration in Eq. (\ref{fml_auxdelta3}),
we replace it with a discrete summation as 
$\int_{\xi_{\rm min}}^{\xi_{\rm max}}d\xi' \delta(\xi' - \xi_{n' k'}) \cdots \approx \sum_i (d \xi)_{i} N_{n'k'}(\xi_{i}) \cdots$. 
The energy grid \{$x_{i}$\} is taken to be non-uniform as elaborated below. 
$N_{n k}(\xi_{i})$ is the integration weight for each point ($nki$),
which is calculated with the following procedure before solving the gap equation: 
(1) Calculate the Kohn-Sham energy eigenvalues on a $k$-point mesh denser than that used for the gap equation, 
(2) apply a tetrahedron-interpolation method to the k-point mesh and evaluate $N_{nk}(\xi_{i})$, 
and (3) calculate optimum $N_{nk}(\xi_{i})$ for the sparse $k$-point grid 
for the gap equation using a reverse interpolatoin method~(Sec.~\ref{app_weight}).

\begin{figure}[bt]
  \includegraphics[width=8cm]{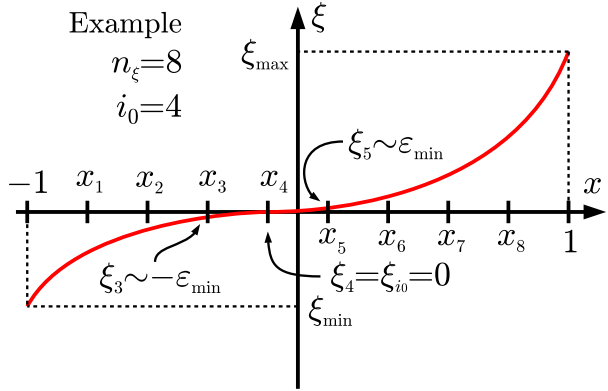}
  \caption{\label{fig_eng_grid}
    (Color online) A schematic illustration of the energy grid.
    It is represented in Eqs. (\ref{fml_energy_grid}) and (\ref{fml_energy_grid_weight});
    we can easily control the accuracy by tuning parameters in those equations.
  }
\end{figure}

We use the following energy grid and the weight of the each point;
\begin{align}
  \xi_{i} &= (x_{i} - x_{i_0}) \varepsilon_{\rm min} \frac{n_{\xi}}{2} \exp
  \left[ a \left( |x_{i}- x_{i_0}| - \frac{2}{n_{\xi}} \right) \right],
  \label{fml_energy_grid}
  \\
  (d \xi)_{i} &= (dx)_{i} (1 + a |x_{i} - x_{i_0}|) \varepsilon_{\rm min} \frac{n_{\xi}}{2} \exp
  \left[ a \left( |x_{i}- x_{i_0}| - \frac{2}{n_{\xi}} \right) \right],
  \label{fml_energy_grid_weight}
\end{align}
where $n_{\xi}$ is the number of energy grid ($i = 1, 2, 3, \cdots, n_{\xi}$),
$x_i$ and $(d x)_{i}$ are the representative point and the weight 
in the Gauss-Legendre quadrature ($-1<x_i<1$).
We choose $i_0$ from $i = 1, 2, \cdots, n_{\xi}$, so that the following factor is minimized:
\begin{align}
  &\left| \xi_{\rm max} -  (1 - x_{i_0}) \varepsilon_{\rm min} \frac{n_{\xi}}{2} \exp
  \left[ a \left(1 - x_{i_0} - \frac{2}{n_{\xi}} \right) \right]\right|
  \nonumber \\  
  +
  &\left| \xi_{\rm min} -  (- 1 - x_{i_0}) \varepsilon_{\rm min} \frac{n_{\xi}}{2} \exp
  \left[ a \left(1 + x_{i_0} - \frac{2}{n_{\xi}} \right) \right]\right|,
\end{align}
where
\begin{align}
  a = \max 
  \left[
    \frac{1}{1 - x_{i_0} - 2 / n_{\xi}} 
    \ln \left( \frac{  \xi_{\rm max}}{(1 - x_{i_0}) \varepsilon_{\rm min} n_{\xi} / 2} \right),
    \frac{1}{1 + x_{i_0} - 2 / n_{\xi}} 
    \ln \left( \frac{- \xi_{\rm min}}{(1 + x_{i_0}) \varepsilon_{\rm min} n_{\xi} / 2} \right)
    \right].
\end{align}
This energy grid has the following properties (see Fig. \ref{fig_eng_grid}):
\begin{enumerate}
  \item It ranges between $\xi_{\rm min}$ and $\xi_{\rm max}$.
  \item The minimum energy scale is $\varepsilon_{\rm min}$.
\end{enumerate}
Then, we can easily control the accuracy by tuning $n_\xi$ and $\varepsilon_{\rm min}$.
%
%

By using auxiliary energy grid, 
we can calculate the quasiparticle density of states (QPDOS) in a superconducting state
as follows:
\begin{align}
  N_{\rm S}(\varepsilon) &= \sum_{n k} \delta(\varepsilon - \varepsilon_{n k})
   = \sum_{n k} \int d \xi \delta(\xi - \xi_{n k}) \delta(\varepsilon - \varepsilon_{n k}(\xi))
   \simeq \sum_{n k i} (d \xi)_{i} N_{n k}(\xi_{i}) \delta(\varepsilon - \varepsilon_{n k}(\xi_{i})).
   \label{fml_sdos}
\end{align}
The four-dimensional (${\bm k}$ and $i$) integration in Eq. (\ref{fml_sdos}) is performed 
by using the pentachoron scheme~(See Sec. \ref{app_4dtetra}).

\subsection{Details of $k$ integrations}

\subsubsection{Reverse interpolation of weight}\label{app_weight}

We consider the $k$-integration as follows: 
\begin{align}
  \langle X \rangle = \sum_{k} X_k w(\varepsilon_k).
\end{align}
If this integration has the following conditions,
it is efficient to interpolate $X_k$ into a denser $k$ grid and 
evaluate that integration in a dense $k$ grid. 
\begin{itemize}
\item $w(\varepsilon_k)$ is sensitive to $\varepsilon_k$
  (e. g. the step function, the delta function, etc.) 
  and requires $\varepsilon_k$ on a dense $k$ grid.
  
\item The numerical cost to obtain $X_k$ is much larger than that to obtain $\varepsilon_k$
  (e. g. the polarization function).
\end{itemize}

This method is performed as follows: 
\begin{enumerate}
\item We calculate $\varepsilon_k$ on a dense $k$grid.
\item We calculate $X_k$ on a coarse $k$ grid and
  obtain that on a dense $k$ grid by using the linear interpolation, 
  the polynomial interpolation, the spline interpolation, etc.
  
  \begin{align}
    X_k^{\rm dense} = \sum_{k'}^{\rm coarse} F_{k k'} X_{k'}^{\rm coarse}
    \label{fml_interpolatedense}
  \end{align}

\item We evaluate that integration in the dense $k$ grid.
  
  \begin{align}
    \langle X \rangle = \sum_{k}^{\rm dense} X_k^{\rm dense} w_k^{\rm dense}
  \end{align}
    
\end{enumerate}

When $X_k$ is a multicomponent array, 
e. g. $X_k = \varphi^*_k(r) \varphi^*_{k+q}(r') \varphi_k(r') \varphi_{k+q}(r)$ for Eq. (\ref{fml_pi0}),
the computational cost for evaluating Eq. (\ref{fml_interpolatedense}) and the memory size for $X_k^{\rm dense}$
become very large.
To avoid this difficulty, 
we developed a method to obtain the result identical to the above result without interpolating $X_k$ into a dense $k$ grid.
Namely, 
we calculate the integration weight on a coarse $k$ grid
from that on a dense $k$ grid; we call it reverse interpolation.
Therefore, if we require 

\begin{align}
  \sum_k^{\rm dense} X_k^{\rm dense} w_k^{\rm dense}
  =
  \sum_k^{\rm coarse} X_k^{\rm coarse} w_k^{\rm coarse},
\end{align}

we obtain

\begin{align}
  w_k^{\rm coarse} = \sum_k^{\rm dense} F_{k ' k} w_{k'}^{\rm dense}.
\end{align}

The numerical procedure for this method is as follows: 
\begin{enumerate}
\item We calculate the integration weight on a dense $k$ grid $w_k^{\rm dense}$
  from $\varepsilon_k$ on that grid.
\item We obtain the integration weight on a coarse $k$ grid ($w_k^{\rm coarse}$)
  by using the reverse interpolation method.
\item We evaluate that integration in a coarse $k$ grid
  where $X_k$ was calculated.
\end{enumerate}
This reverse interpolation method is employed in evaluating Eqs. (\ref{fml_pi0}),
(\ref{fml_auxdelta3}), and (\ref{fml_sdos}).

\subsubsection{Four dimensional numerical integration scheme for DOS}\label{app_4dtetra}

For evaluating accurately the four-dimensional integration in Eq. (\ref{fml_sdos}),
we construct a method by extending the tetrahedron method to the four-dimensional case.
We consider the following integration:
\begin{align}
  \langle X \rangle = \int_{\rm BZ} d^3 k \int d \xi X_k(\xi) \delta(\varepsilon - \varepsilon_{k}(\xi)),
\label{fml_4dintegrate}
\end{align}
where $X_k(\xi)$ and $\varepsilon_{k}(\xi)$ are smooth functions of $k$ and $\xi$;
in the calculation of the QPDOS, $X_k(\xi) = N_{n k}(\xi)$ and 
$\varepsilon_{k}(\xi) = \sqrt{\Delta^2_{n k}(\xi)+\xi^2}$.

We divide four-dimensional $(k,\xi)$ space into $24 \times N_k \times (N_{\xi} - 1)$ pentachora.
If we assume $X_{k}(\xi)$ and $\varepsilon_{k}(\xi)$ as linear functions
of $k$ and $\xi$ in each pentachoron,
we can obtain the following result of Eq. (\ref{fml_4dintegrate}) in a pentachoron.
\begin{align}
  \langle X \rangle_P \approx 
  \frac{S_{\varepsilon_{k}(\xi) = \varepsilon}}{|\nabla_{k,\xi} \varepsilon_{k}(\xi)|}
  \langle X_{k}(\xi) \rangle_{\varepsilon_{n k}(\xi) = \varepsilon}
  \equiv \sum_{i=1}^{5} w_i X_i,
  \label{fml_pentachron}
\end{align}
where $S_{\varepsilon_{k}(\xi) = \varepsilon}$ is the volume of the region
in which $\varepsilon_{k}(\xi)$ becomes $\varepsilon$, 
$\langle X_{k}(\xi) \rangle_{\varepsilon_{k}(\xi) = \varepsilon}$
indicates $X_{k}(\xi)$ averaged in that region,
$X_i$ is $X_{k}(\xi)$ at the each corner of the pentachoron;
$w_i$ can be calculated analytically from $\varepsilon_k(\xi)$ (See  App. \ref{app_pentachron}).

%
%
\section{Results}

In this section, 
we show our results of YNi$_2$B$_2$C:
the normal-state band structure, Fermi surfaces, phonon dispersion, superconducting transition temperature, 
gap functions, and quasiparticle DOS in the superconducting phase.
We used the DFT code Quantum ESPRESSO\cite{QE-2009}, 
which employs plane-waves and the pseudopotential to describe 
the Kohn-Sham orbitals and the crystalline potential, respectively.
We obtain phonon frequencies and electron-phonon vertices by using 
density functional perturbation theory (DFPT)\cite{RevModPhys.73.515}.
We employ the optimized tetrahedron method\cite{PhysRevB.89.094515, dfpt_tetra} 
for the Brillouin zone integrations
in calculations of the charge density, phonons, and the polarization function.
We used our open-source program {\sc superconducting toolkit}
\cite{SuperconductingToolkit} for the calculations concerning SCDFT.

\subsection{Electronic structures of normal state}

The calculations were done with the GGA-PBE\cite{PhysRevLett.78.1396} exchange-correlation functional.
We set the plane-wave cutoff for the Kohn-Sham orbitals to 50 Ry.
We used the ultrasoft pseudopotentials \cite{PhysRevB.41.7892} in Ref. \onlinecite{PseudoLib_yni2b2c}.
We also performed the calculations with the LDA-PZ functional \cite{PhysRevB.23.5048}
and refer to them for comparison when necessary.
The numerical conditions are summarized in Table \ref{tbl_numcond}.
We performed calculations with $4^3$, $6^3$, and $8^3$ q-point grid and
obtained the converged result with the $6^3$ grid.
\begin{table}[!b]
  \begin{center}
    \caption{\label{tbl_numcond} Numerical conditions. 
      For the definitions of $n_{\xi}$ and $\varepsilon_{\rm min}$, see Sec.~\ref{sec:anxiliary-gap}.
    }
    \begin{tabular}{cc}
      \hline
      $k$ grid (structure and charge density optimization)     & $12 \times 12 \times 12$ \\
      $q$ grid (wavenumber of phonons)      & $6 \times 6 \times 6$ \\
      $k$ grid (density of states)          & $40 \times 40 \times 40$ \\
      The number of bands (gap equation) & 50 bands \\
      The number of bands (polarization function $\Pi_0$) & 50 bands \\
      The number of points for energy grid $n_{\xi}$ & 100 \\
      $\varepsilon_{min}$ in energy grid & $10^{-6}$ Ry \\
      \hline
    \end{tabular}
  \end{center}
\end{table}
\begin{table}[!b]
  \begin{center}
    \caption{\label{tbl_lattice_parameter} 
      Results of the structure optimization in comparison with experimental data\cite{Siegrist1994135}.
      Crystal structure is depicted in Fig. \ref{fig_crys_kpath} (a).
    }
    \begin{tabular}{cc}
      \hline
      lattice constant $a$ [\AA] & 3.48 (LDA) /  3.51 (GGA)  /  3.533 (Experiment) \\
      lattice constant $c$ [\AA] & 10.19 (LDA) / 10.31 (GGA)  / 10.566 (Experiment) \\
      B-C length      [\AA] & 1.483 (LDA) /  1.494 (GGA)  /  1.492 (Experiment) \\
      \hline
    \end{tabular}
  \end{center}
\end{table}
\begin{figure}[!b]
  \includegraphics[width=8cm]{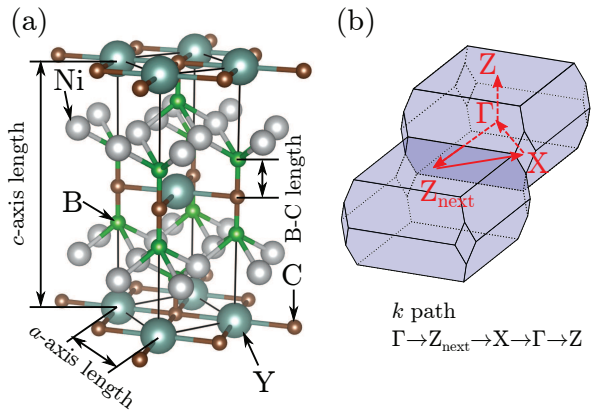}
  \caption{(Color online) (a) Crystalline structure of YNi$_2$B$_2$C.
  (b) Brillouin zone and $k$ path}
  \label{fig_crys_kpath}
\end{figure}

First we performed the structure optimization 
(crystalline structure is depicted in Fig. \ref{fig_crys_kpath} (a));
the optimized and experimental structural parameters are given in Table \ref{tbl_lattice_parameter}.
The parameter $c$ is underestimated by 2\%. Similar underestimation can be seen in a previous report~\cite{PhysRevB.89.104503} and this is probably due to the drawback with the GGA-PBE functional~\cite{PhysRevB.89.104503}.
The later calculations were based on the theoretically optimized lattice parameters, though we have found that the setting of the parameters (either theoretically optimized or experimentally observed values) yields little difference in the calculated phononic and superconducting properties.

%
\begin{figure}[!bt]
  \includegraphics[width=15.0cm]{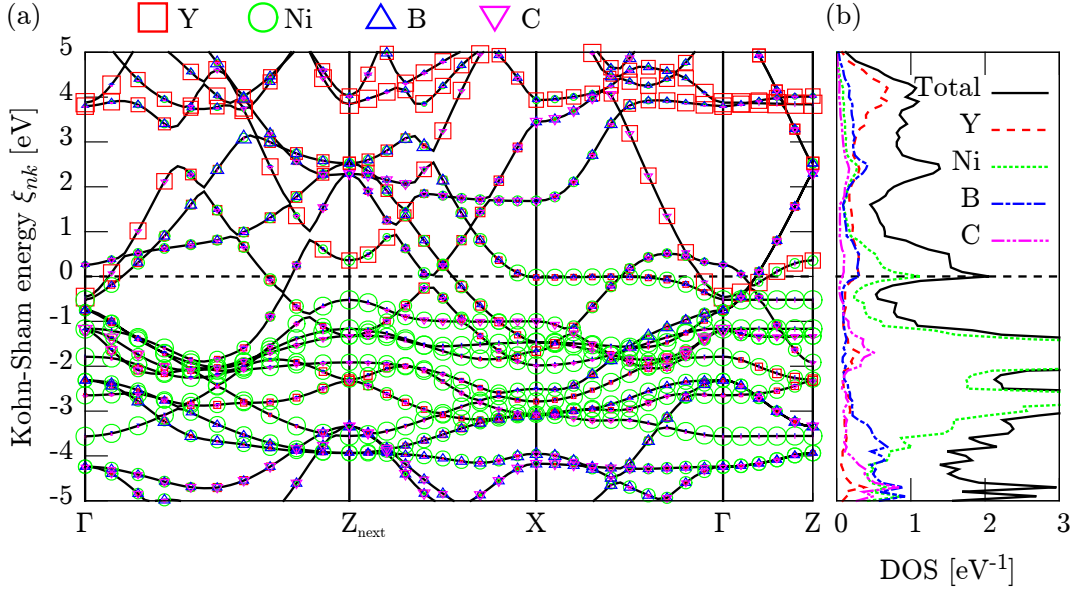}
  \caption{(Color online) (a) Band structure of YNi$_2$B$_2$C. 
    Sizes of red squares, green circles,
    blue upward triangles, and magenta downward triangles indicate
    the amount of components of atomic orbitals of Y 4{\it d}, Ni 3{\it d}, 
    B 2{\it s}2{\it p}, and C 2{\it s}2{\it p}, 
    respectively.
    (b) Partial- and total- density of states.
    The black solid line, the red dashed line, the green doted line, 
    the blue dashed-doted line, and the magenta dashed-two doted line
    indicate the total-, Y 4$d$, Ni 3$d$, B 2$s$2$p$, and C 2$s$2$p$ DOS, respectively.
  }
  \label{fig_band}
\end{figure}
Figure \ref{fig_band} (a) shows the calculated band structure of YNi$_2$B$_2$C
[the $k$ path is depicted in Fig. \ref{fig_crys_kpath} (b)].
We here describe the contributions of the atomic orbitals--Y 4$d$, Ni 3$d$, B 2$s$2$p$, and C 2$s$2$p$--
as the size of the symbols; for example, the Ni 3$d$ contribution to the Kohn-Sham state $nk$ $p^{{\rm Ni} 3 d}_{n k}$ is defined by
\begin{align}
  p^{{\rm Ni} 3 d}_{n k} = \sum_{\tau = {\rm Ni} 1, {\rm Ni} 2} \sum_{m} 
  |\Braket{\varphi_{\tau d m} | \varphi_{n k}}|^2 .
  \label{fml_projection}
\end{align}
We also depict the total- and the partial- density of states 
in Fig \ref{fig_band} (b).
This band structure agrees with the one obtained in the previous study with a GGA functional\cite{PhysRevB.67.104507}.
%
There is a flat band near the Fermi level on the $X-\Gamma$ line; 
electronic states in this flat band
consist mainly of Ni 3{\it d} state.
The total density of states at the Fermi level is 29 states per Ry, spin, and unit cell, to which
Y 4{\it d}, Ni 3{\it d}, B 2{\it s}2{\it p}, and C 2{\it s}2{\it p} states contribute by
16.5\%, 62.7\%, 16.6\%, and 4.2\%, respectively.
The large contribution from the Ni 3{\it d} orbital mainly
comes from the proximity of the flat band on the $X-\Gamma$ line.

\begin{figure}[!bt]
  \includegraphics[width=15.0cm]{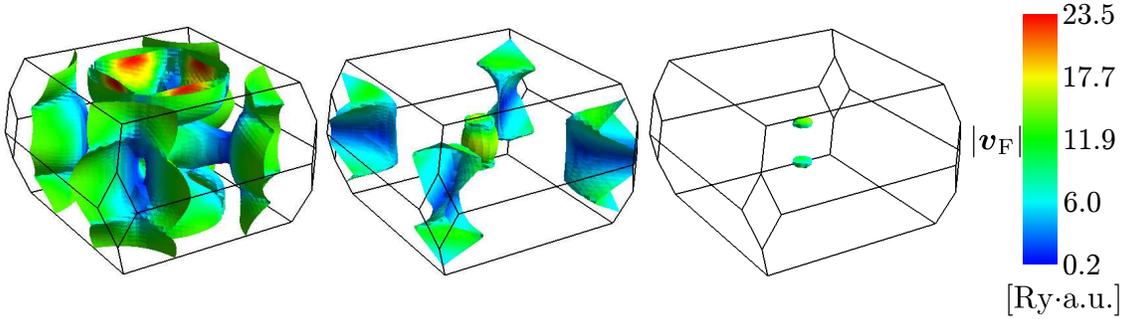}
  \caption{(Color online) Fermi velocity of the electronic states on Fermi surfaces.
    The red, green, and blue region have a high, middle, and low Fermi velocity, respectively.
    Fermi surfaces in this paper are drawed by using FermiSurfer\cite{fermisurfer} program
    which is developed by us.
    \label{fig_vfermi_noshift}}
\end{figure}

\begin{figure}[!bt]
  \includegraphics[width=15.0cm]{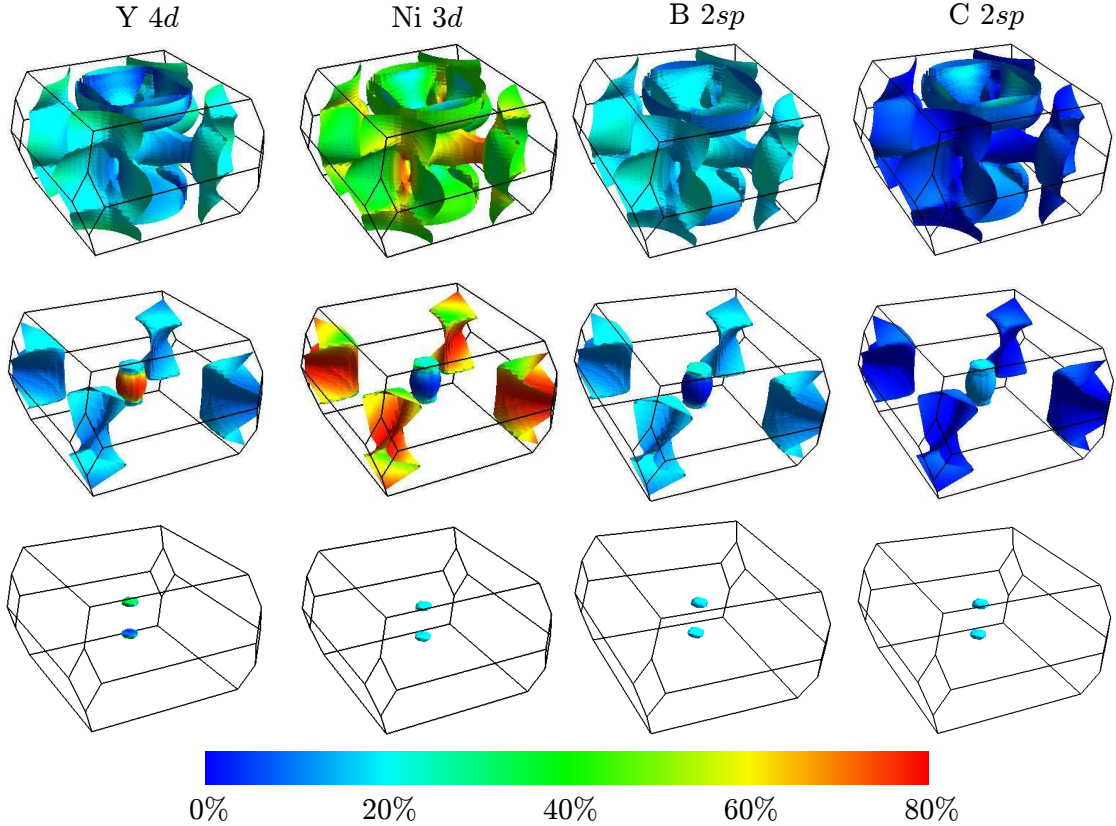}
  \caption{(Color online) Projection of atomic orbitals 
    Y 4{\it d}, Ni 3{\it d}, B 2{\it s}2{\it p}, and C 2{\it s}2{\it s}
    on Fermi surfaces 
    ($|\Braket{\varphi_{\rm Atom} | \varphi_{n k}}|^2$).
  }
  \label{fig_proj_noshift}
\end{figure}

Figure \ref{fig_vfermi_noshift} shows the Fermi surfaces, 
on which we describe the distribution of the Fermi velocity with a color plot.
It varies largely over Fermi surfaces;
the ratio of its maximum to minimum is about 100.
We calculate the projections of the atomic orbitals
Y 4{\it d}, Ni 3{\it d}, B 2{\it s}2{\it p}, and C 2{\it s}2{\it s}, 
to the electronic states on Fermi surfaces (Fig. \ref{fig_proj_noshift}).
There is no regions dominated by B 2{\it s}2{\it p}, and C 2{\it s}2{\it p} orbitals.
Comparing Fig. \ref{fig_vfermi_noshift} and Fig. \ref{fig_proj_noshift},
we found that the Fermi velocity is particularly small in the regions
where Ni $3 d$ orbitals are dominant.

\subsection{Phonons and electron-phonon interactions}

We next calculated the phonon and electron-phonon interaction.
The calculated frequencies of the Raman-active modes are given in Table \ref{tbl_raman}.
Results from the previous Raman scattering experiment and first-principles calculation with the all-electron 
full potential linear augmented plane wave (FLAPW) method and the GGA-PBE functional
are also shown.
Our results show good agreement with both previous experimental and theoretical results.
\begin{table}[!bt]
  \begin{center}
    \caption{\label{tbl_raman} 
      Calculated Raman-active phonon frequencies in cm$^{-1}$
      in comparison with those obtained in the previous theoretical work\cite{PhysRevB.67.104507} 
      with full potential linear augmented plane wave (FLAPW) method and the GGA-PBE functional
      and experimental Raman-scattering measurements.}
    \begin{tabular}{cccc}
      \hline
      & This work & Previous (FLAPW)\cite{PhysRevB.67.104507} & Previous (experiment) \\
      \hline
      Ni-$B_{\rm 1 g}$ & 193 & 200 & 199\cite{JLowTempPhys.105.1629}, 198\cite{PhysRevB.50.16726}, 
      193\cite{PhysRevB.53.2237} \\
      Ni-$E_{\rm g}$   & 279 & 271 & 287\cite{JLowTempPhys.105.1629}, 282 \cite{PhysRevB.50.16726}\\
      B-$E_{\rm g}$   & 461 & 447 & 460\cite{JLowTempPhys.105.1629}, 470\cite{PhysRevB.50.16726} \\
      B-$A_{\rm 1g}$  & 836 & 821 & 813\cite{JLowTempPhys.105.1629}, 832\cite{PhysRevB.50.16726},
      823\cite{PhysRevB.53.2237}, 847\cite{PhysRevB.52.6208}  \\ 
      \hline
    \end{tabular}
  \end{center}
\end{table}

\begin{figure}[!bt]
  \includegraphics[width=8cm]{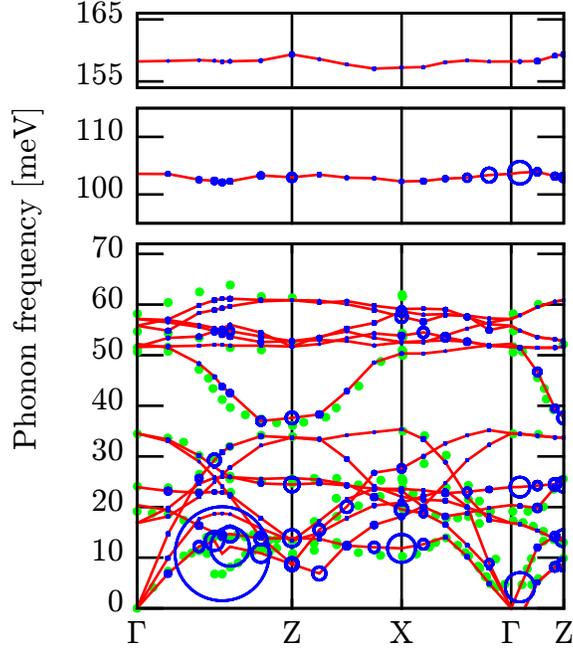}
  \caption{(Color online) Phonon dispersion. The radii of circles indicate
    magnitude of $\lambda_{q \nu}$.
    Green filled circles indicate 
    results of the neutron diffraction \cite{PhysRevB.89.104503}. }
  \label{fig_disp_lambda}
\end{figure}

\begin{figure}[!bt]
  \includegraphics[width=8cm]{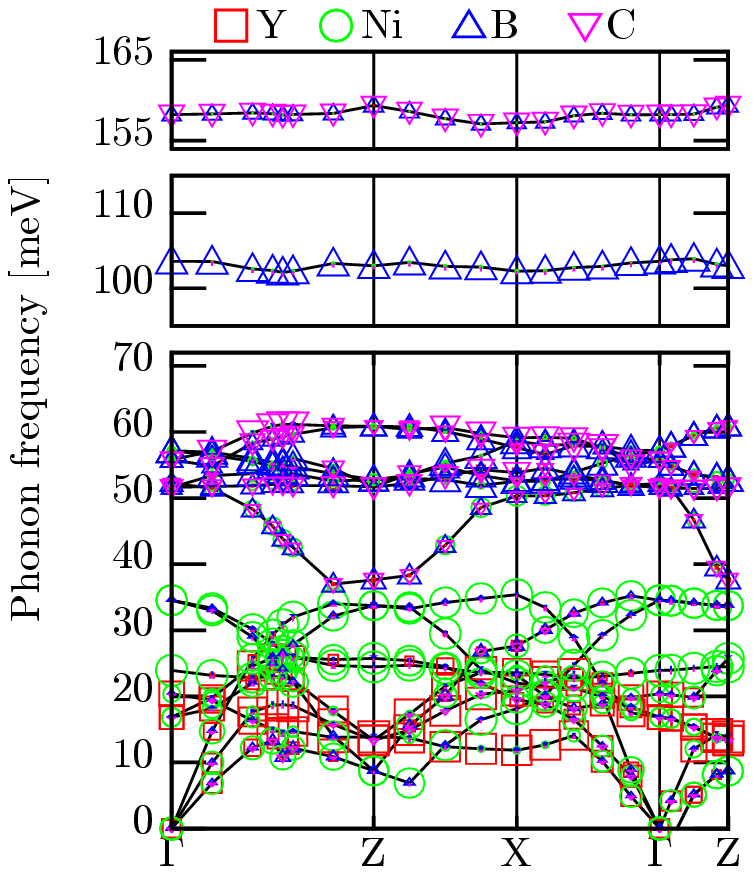}
  \caption{(Color online) Phonon dispersion. Sizes of red squares, green circles,
    blue upward triangles, and magenta downward triangles indicate 
    magnitude of components of Y, Ni, B, and C, respectively
    of the displacement pattern.}    
  \label{fig_disp_atom}
\end{figure}

We show the calculated phonon dispersions in Fig. \ref{fig_disp_lambda}.
The whole spectra agree well with those obtained with the neutron scattering measurement \cite{PhysRevB.89.104503} except for the behavior of the TA band around $q \sim 0.55 {\rm \Gamma Z_{next}}$;
although this mode shows strong softening in experiments,
the softening obtained in our calculation is not as strong. %
We observe imaginary modes in the vicinity of the $\Gamma$ point along the $\Gamma$--Z line;
this indicates that the system theoretically favors long-period modulation
though such a structure has not clearly been observed experimentally.
Assuming that the present imaginary modes is an artifact of the present approximation,
we just neglect them because phonons with such long wavelength 
do not affect the superconductivity.
We also depict the electron-phonon coupling constant 
\begin{align}
  \label{fml_lambdaq}
  \lambda_{q \nu} = \sum_{k n n'}  \frac{2}{N(\varepsilon_{\rm F}) \omega_{q \nu}}
  |g_{n' k + q n k}^{\nu}|^2 
  \delta(\xi_{n k}) \delta(\xi_{n' k+q})
\end{align}
of each phonons (the branch dependent Fr\"ohlich parameter) as radii of circles,
where $N(\varepsilon_{\rm F})$ is the density of states at the Fermi level;
the TA mode has large electron-phonon interaction.
The contribution of each atom to each phonon mode can be seen in Fig. \ref{fig_disp_atom};
there are roughly six groups in this phonon dispersion such as
three acoustic branches ranging from 0 meV to 30 meV,
Y-dominant branches ranging from 10 meV to 25 meV,
Ni-dominant branches ranging from 20 meV to 35 meV,
B-C branches ranging from 35 meV to 60 meV,
B-dominant branch at approximately 102 meV,
and
B-C branch at approximately 159 meV.
Non-dispersive branches at 102 meV and 159 meV have been
observed by the time-of-flight neutron spectroscopy experiment \cite{PhysRevLett.109.057001}
in good agreement with our calculation.

\begin{figure}[!bt]
  \includegraphics[width=15.0cm]{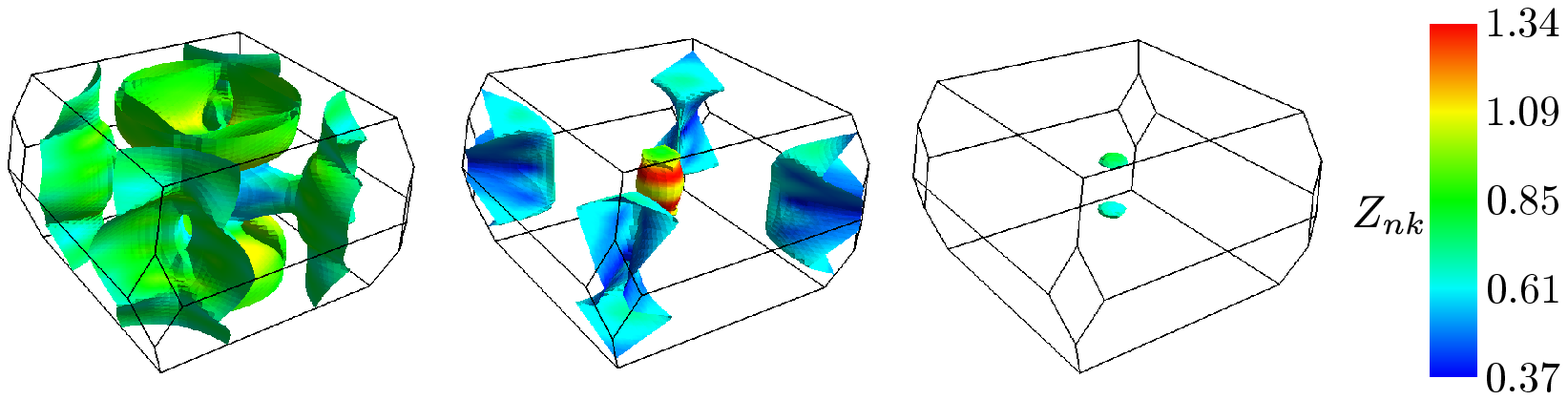}
  \caption{(Color online) Electron-phonon renormalization $Z_{n k}$ 
    on Fermi surfaces. \label{fig_z_noshift} 
  }
\end{figure}

The electron-phonon renormalization $Z_{n k} \equiv Z(\{g_{n' k' n k}\}, \{\omega_{k'-k}\}, \xi_{n k})$
of electronic states on the Fermi surfaces
are shown in Fig. \ref{fig_z_noshift}.
This has large anisotropy and the ratio between the maximum
and the minimum of the $Z_{n k}$ is approximately 4;
this ratio is close to the value previously determined 
with the dHvA experiment \cite{TERASHIMA1995459}
referring to the band structure calculation \cite{Yamauchi2004225}.
Comparing Fig. \ref{fig_proj_noshift} and Fig. \ref{fig_z_noshift},
we can see that the electronic states that have small $Z_{n k}$
consist mainly of Ni 3{\it d} orbitals.

\begin{figure}[!bt]
  \includegraphics[width=8cm]{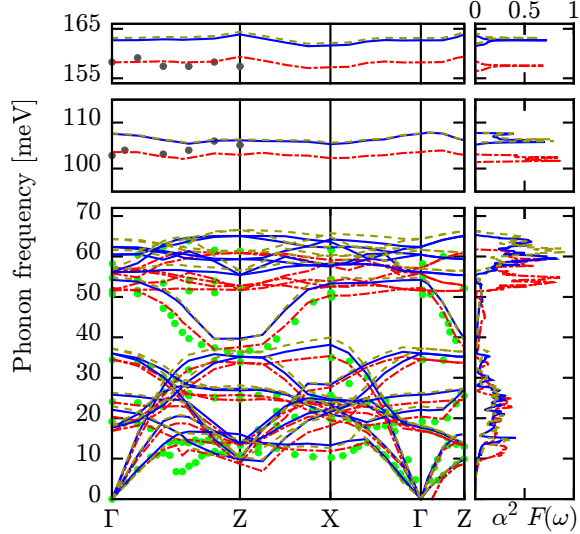}
  \caption{(Color online) Phonon dispersions and Eliashberg functions
    computed in three different conditions.
    The red dashed-doted line, the blue dolid line, and the yellow dashed  line
    indicate those computed by using
    the GGA functional with the GGA geometry (a geometry optimized with the GGA functional), 
    the LDA functional with the LDA geometry, and
    the GGA functional with the LDA geometry, respectively.
    Green and gray filled circles indicate 
    results of the neutron diffraction experiment in Ref. \onlinecite{PhysRevB.89.104503}
    and Ref. \onlinecite{PhysRevLett.109.057001}, respectively. }
  \label{fig_disp2}
\end{figure}

From the branch dependent Fr\"ohlich parameter $\lambda_{q \nu}$, 
we compute the total Fr\"ohlich parameter and the averaged phonon frequency,
\begin{align}
  \lambda = \sum_{q \nu} \lambda_{q \nu},
  \qquad
  \omega_{\rm ln} = \exp \left[ \frac{1}{\lambda} \sum_{q \nu} 
   \ln(\omega_{q \nu}) \lambda_{q \nu} \right].
\end{align}
We obtain $\lambda = 0.72$, and $\omega_{\rm ln} = 270$ K (23.3 meV) by using the GGA-PBE functional;
we obtain $\lambda = 0.54$, and $\omega_{\rm ln} = 291$ K (25.1 meV) by using the LDA-PZ functional.
We can find the origin of the functional dependence of
the phonon dispersion and the as Fr\"ohlich parameter follows.
Figure \ref{fig_disp2} shows phonon dispersions and Eliashberg functions
computed in three different conditions, namely,
the GGA functional with the GGA geometry (a geometry optimized with the GGA functional), 
the LDA functional with the LDA geometry, and
the GGA functional with the LDA geometry.
When we use the LDA geometry, the phonon is hardened
because of the underestimated interatomic distance.
From this hardened phonon, we obtain a small $\lambda$.
This overestimation of the phonon frequency is improved by using the GGA geometry.
We see below that 
this dependence on the exchange-correlation functional yield some variation of the resulting $T_{\rm c}$,
though the superconducting solution is robustly present.
The Fr\"ohlich parameter computed with the GGA functional is slightly smaller
than that from the specific heat measurement
$\lambda_{\rm S-H}=\gamma_{\rm exp}/\gamma_{\rm band}-1=0.82$,
where $\gamma_{\rm exp}=18.2$ mJ$/$mol$/$K$^2$ is
the Sommerfeld parameter from the specific heat measurement\cite{PhysRevB.52.16165}
and
$\gamma_{\rm band}=10.0$ mJ$/$mol$/$K$^2$ is that parameter obtaind from the band structure
computed in the current work.
This underestimation probably comes from the incomplete reproduction of the phonon softenning
of the TA band around $q \sim 0.55 {\rm \Gamma Z_{next}}$ (See Fig. \ref{fig_disp_lambda}).

\subsection{Superconducting gaps and transition temperature}

\begin{figure}[!bt]
  \begin{center}
    \includegraphics[width=8.0cm]{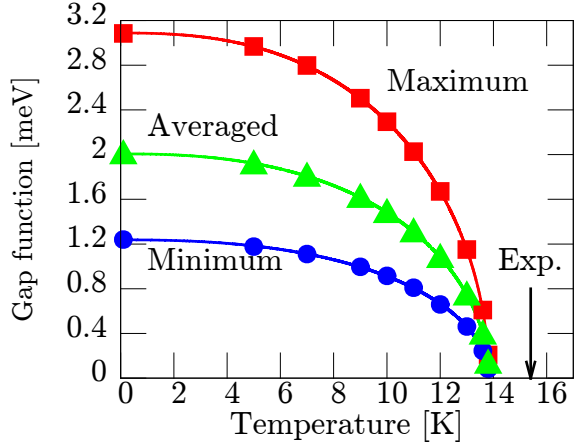}
    \caption{(Color online) Calculated and experimental superconducting transition temperature.
      Red squares, green triangles, and blue circles indicate the maximum-, the averaged-, 
      and the minimum superconducting gaps on Fermi surfaces;
      solid lines are a fit of these gaps with a function
    $\Delta(T) = \Delta_{0} \{1-(T/T_{\rm c})^p\}^{1/q}$ via $\Delta_0, T_{\rm c}, p$, and $q$.}
    \label{fig_tc_noshift}
  \end{center}
\end{figure}

Let us now move on to the superconducting properties.
We calculated the superconducting gap function at various temperatures. The values of the gap function averaged over the Fermi surfaces for the respective temperatures, as well as the maximum and minimum values are plotted in Fig. \ref{fig_tc_noshift}.
The calculated transition temperature where superconducting gaps disappear, 13.8~K, 
agrees well with the experimental value, 15.4~K.
We also obtained the superconducting solution with the LDA-PZ functional; although the resulting $T_{\rm c}$ is 8.73~K, this result indicates that the superconducting phase is numerically robust against the change of the exchange-correlation functional.
The calculated isotope effect exponent for boron atoms $\alpha_{\rm B}$ is 0.16, 
in fair agreement with the experimentally observed values 
($\alpha_{\rm B}=0.11 \pm 0.05$, $0.21 \pm 0.07$\cite{Cheon199935}, 
$0.25\pm0.04$, $0.27\pm0.07$, $0.26\pm0.03$\cite{Lawrie1995159}).

\begin{figure}[!bt]
  \includegraphics[width=15.0cm]{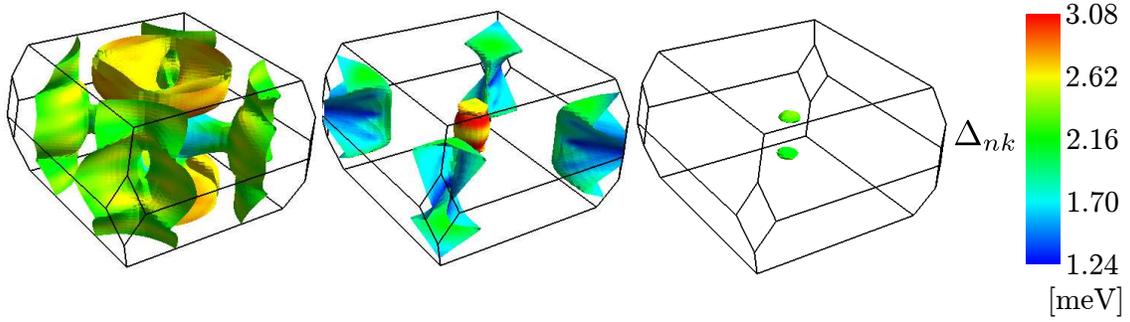}
  \caption{(Color online) Superconducting gap functions $\Delta_{n k}$ on Fermi surfaces at 0.1 K.
  }
  \label{fig_delta_noshift}
\end{figure}

\begin{figure}[!bt]
  \includegraphics[width=15.0cm]{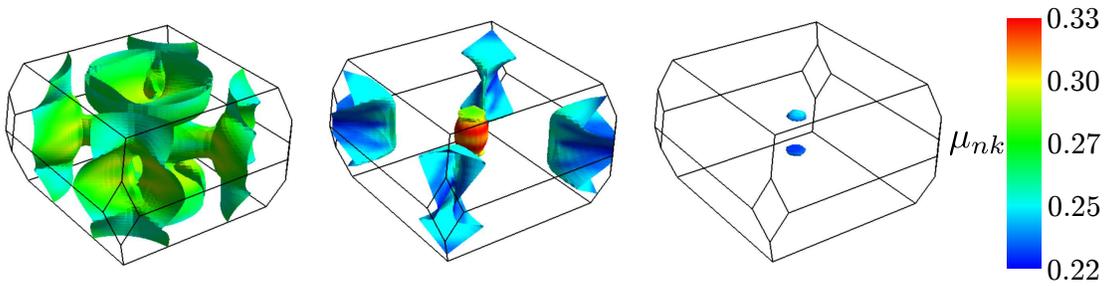}
  \caption{(Color online) The $nk$ dependent Coulomb potential defined in Eq. (\ref{fml_mu_k}).
  }
  \label{fig_mu_k}
\end{figure}

We depict the superconducting gap function $\Delta_{n k}$ on Fermi surfaces at low temperature (0.1 K)
in Fig. \ref{fig_delta_noshift}.
\footnote{Calculation at the 0 K needs special treatment, and the result at 0 K
and that at 0.1 K are almost the same; 
therefore we calculate at 0.1 K instead of the 0 K.}
As we expected, superconducting gaps of YNi$_2$B$_2$C is anisotropic;
similar to the case of $Z_{n k}$, electronic states that have a small
superconducting gap consist of Ni 3{\it d} orbitals.
However, the degree of anisotropy is smaller than that of 
the electron-phonon coupling;
the ratio between the maximum and minimum of the gap functions on Fermi surfaces is 2.4.
This suppression of anisotropy comes from the following two reasons:
First, the $nk$ dependence of the screened Coulomb interaction
cancels the $nk$-dependent pairing induced by the phonon.
Figure \ref{fig_mu_k} shows the $nk$ dependent Coulomb potential
\begin{align}
  \mu_{n k} \equiv \sum_{q n'} \delta(\xi_{n' k+q}) K^{\rm ee}[V_{n' k+q n k}](\xi_{n k}, \xi_{n' k+q}).
  \label{fml_mu_k}
\end{align}
The sign of the Coulomb repulsion and the phonon mediated attraction is opposite,
while their absolute values are positively correlated. 
The dependence of their sum is thus moderated.
Second, the integration kernel in Eq. (\ref{fml_ksgapeq}) is reduced by a factor $1/(1+Z_{nk})$;
this renormalization is large in the region where 
the electron-phonon interaction is strong. 
Consequently, the anisotropy of the integration kernel
becomes smaller than that of the electron-phonon interaction,
and the anisotropy of the superconducting gap is suppressed.

To examine the effect of the exchange-correlation kernel $\delta^2 E_{\rm XC}/\delta^2 \rho$ in
the electron-electron kernel [Eq. (\ref{fml_screenedcoulomb})], 
we calculate superconducting gap by using RPA also;
the difference in the $T_{\rm c}$ was less than 0.1~K compared with that from the ALDA.
Therefore, in YNi$_2$B$_2$C, the effect of the exchange-correlation kernel 
is very small at the ALDA level.
We perform a converse calculation of the Coulomb pseudopotential $\mu^*$, 
which is usually treated as a fitting parameter for McMillan's formula
\cite{PhysRev.167.331,Dynes1972615}
\begin{align}
  \label{mcmillan}
  T_{\rm c} = \frac{\omega_{\ln}}{1.2} \exp \left(\frac{-1.04 (1+\lambda)}
  {\lambda - \mu^{*}(1+0.62 \lambda)}\right).
\end{align}
Namely, we determined $\mu^*$ so that the transition temperatures calculated with the RPA-SCDFT and ALDA-SCDFT 
are reproduced with $\lambda = 0.72$ and $\omega_{\ln} = 267$ K;
We obtain $\mu^*=0.053$ in the both cases. Notably, this value is far smaller than the conventional range 
(0.10--0.13 \cite{PhysRev.125.1263}). 
This indicates that the $nk$-averaging approximation, which is applicable to ordinary materials, 
substantially underestimate $T_{\rm c}$ and the anisotropy is important for the observed high $T_{\rm c}$.

\begin{figure}[!bt]
  \includegraphics[width=8cm]{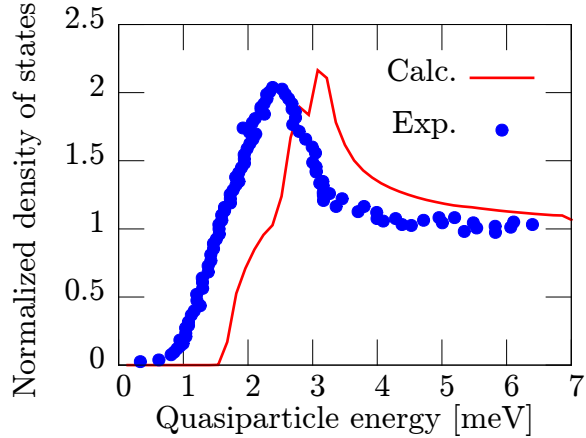}
  \caption{(Color online) Calculated superconducting quasiparticle density of states at 0.1 K and
    experimental tunnel conductance spectrum at 0.5 K \cite{PhysRevB.67.014526}.}
  \label{fig_qpdos_noshift}
\end{figure}

%
Using the calculated $k$-dependent gap function, 
we next evaluated the quasiparticle density of states (QPDOS) in the superconducting phase. 
The calculated QPDOS is compared to the density of states extracted from the tunnel-conductance measurement
\cite{PhysRevB.67.014526} (Fig. \ref{fig_qpdos_noshift});
although there is a visible discrepancy between the peak positions of the calculated QPDOS 
and the experimental spectrum, their whole shapes are very similar (Fig. \ref{fig_tc_noshift}).
Note that we did not use the smearing method for the four-dimensional integral;
therefore, the broadened peak structure  definitely originates from the $k$-space variation of the gap function.
If we use the smearing, we can not distinguish a broadened peak
made by the variation of the gap function and one made by the smearing.
\section{Discussion}

As revealed with Fig. \ref{fig_delta_noshift}, 
  the $k$-space distribution shows full $s$-wave gap
with subtle dependence which is obviously composed of multiple
high order spherical harmonics.
As a result, $\Delta_{nk}$ shows continuous spectrum across the multiple
Fermi surfaces; namely, multiband extended s-wave gap.
Here we note
that the superconducting gap has significant intra-band anisotropy;
this is in stark contrast with the ``anisotropic gap'' in MgB$_2$, 
where the gap value varies with bands but does not vary much within each Fermi surface\cite{PhysRevB.66.020513}.
%
%
We have found 
a significant correlation between the anisotropy of the superconducting gaps in YNi$_2$B$_2$C and
the variation of the ratio of atomic orbitals on the Fermi surfaces.
The electronic states on the Fermi surfaces in YNi$_2$B$_2$C consist of 
Y 4{\it d}, Ni 3{\it d}, B 2{\it s}2{\it s}, and C 2{\it s}2{\it p};
in particular, the electronic states dominated by Ni 3{\it d} orbitals 
couple to phonons very weakly, consequently exhibiting very small gap.
To evaluate contributions from each atomic orbitals to the superconducting gap, 
we defined the superconducting gaps of each orbitals
$\Delta_{o}$ ($o={\rm Y} 4d, {\rm Ni} 3d, {\rm B} 2s2p, $ and ${\rm C} 2s2p$)
as the fitting parameters of $\Delta_{n k}$ in the following form:
\begin{align}
  \Delta_{n k}^{\rm test} = \Delta_{{\rm Y} 4 d} p^{{\rm Y} 4 d}_{n k} + \Delta_{{\rm Ni} 3d} p^{{\rm Ni} 3 d}_{n k}
  + \Delta_{{\rm B} 2 s 2 p} p^{{\rm B} 4d}_{n k} + \Delta_{{\rm C} 2 s 2 p} p^{{\rm C} 2s 2p}_{n k},
\end{align}
The factors $p_{n k}$s are contributions of the respective atomic orbitals
to the electronic state $\varphi_{n k}$ [Eq. (\ref{fml_projection})]. 
We determined 
$\Delta_{{\rm Y} 4 d}$, $\Delta_{{\rm Ni} 3d}$, $\Delta_{{\rm B} 2 s 2 p}$, and $\Delta_{{\rm C} 2 s 2 p}$
so that the following variance is minimized
\begin{align}
  \sigma^2 = \sum_{n k} \delta(\xi_{n k}) (\Delta_{n k} - \Delta_{n k}^{\rm test})^2;
\end{align}
we obtained $\Delta_{{\rm Y} 4 d} = 2.0$, $\Delta_{{\rm Ni} 3 d} = 1.5$, 
$\Delta_{{\rm  B} 2 s 2 p} = 3.9$, and $\Delta_{{\rm C} 2 s 2 p} = 10.8$, with
the fitting error 
\begin{align}
  \left \langle \frac{\delta \Delta}{\Delta} \right \rangle = 
  \frac{\sum_{n k} \delta(\xi_{n k}) |\Delta_{n k} - \Delta_{n k}^{\rm test}|/|\Delta_{n k}|}
       {\sum_{n k} \delta(\xi_{n k})}
\end{align}
being 12.6 \%.
We also applied a similar analysis on the electron-phonon renormalization $Z_{nk}$:
Namely, 
we fit the electron-phonon renormalization $Z_{n k}$ into the
orbital-dependent form
\begin{align}
  Z_{n k}^{\rm test} = Z_{{\rm Y} 4 d} p^{{\rm Y} 4 d}_{n k} + Z_{{\rm Ni} 3d} p^{{\rm Ni} 3 d}_{n k}
  + Z_{{\rm B} 2 s 2 p} p^{{\rm B} 4d}_{n k} + Z_{{\rm C} 2 s 2 p} p^{{\rm C} 2s 2p}_{n k}.
\end{align}
We obtain $Z_{{\rm Y} 4 d} = 0.85$, $Z_{{\rm Ni} 3 d} = 0.45$, 
$Z_{{\rm  B} 2 s 2 p} = 1.21$, and $Z_{{\rm C} 2 s 2 p} = 4.22$, with
the fitting error $\langle \delta Z/Z \rangle=14.9$.
%
The small value of $Z_{{\rm Ni} 3d}$ indicates that the mixing of Ni 3{\it d} orbitals 
weakens the interaction with the phonons, 
which is the key factor behind the mechanism of the anisotropic gap.

The relatively accurate fitting errors in the above analysis suggest that,
in the real space, the coupling to phonons and gaps at the respective atoms possibly exhibit specific values.
The gap structure varying within the
respective Fermi-surface sheets
is then interpreted to originate from the complicated hybridization
between the atomic orbitals.
A recently developed real-space method\cite{PhysRevLett.115.097002}
could be helpful to substantiate this scenario.

Here we discuss why the Ni 3{\it d} orbital results in the weak electron-phonon interaction.
We infer that the localized nature of Ni 3d orbitals has a crucial role;
this localization affects the screened electron-phonon interaction through the following two possible routes:
It makes the electronic states more sensitive to the deformation potential of the Ni ion, 
which should yield stronger electron-phonon coupling.
On the other hand, the highly localized Ni 3{\it d} electrons participate 
in the local screening of the deformation potential, 
which should make the electron-phonon coupling weak.
In the present case, the latter is considered to be dominant.
%
To confirm this point, we calculate the renormalization factor $Z_{n k}$ 
by using the {\it bare} electron-phonon vertex (note that the electron-phonon vertex employed for the superconducting calculations are usually calculated with the {\it screened} perturbation potential of atomic displacements).
\begin{figure}[!bt]
  \includegraphics[width=15.0cm]{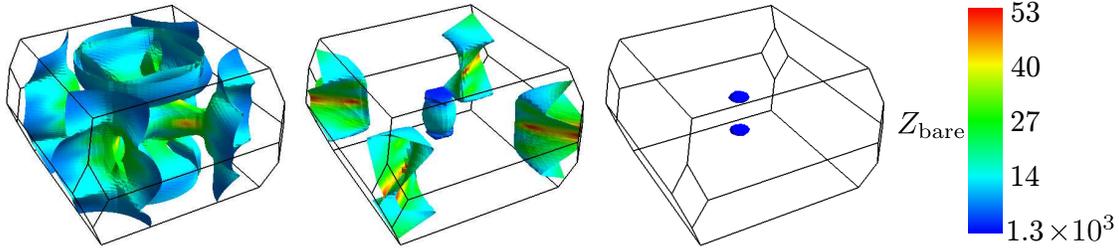}
  \caption{The electron-phonon renormalization $Z$ 
     calculated by using the bare electron-phonon vertex on Fermi surfaces. \label{fig_z_bare}}
\end{figure}
Figure \ref{fig_z_bare} shows the resulting $Z_{\rm bare}$;
Performing the same fitting as before, we obtain 
$Z_{{\rm Y} 4 d} = 6.3 \times 10^3$, $Z_{{\rm Ni} 3 d} = 3.8 \times10^4$,
$Z_{{\rm  B} 2 s 2 p} = -1.3 \times 10^4$, and $Z_{{\rm C} 2 s 2 p} = 6.7 \times 10^3$;
the fitting error is 39.1 \%
\footnote{
The cause of the increase in fitting error is that 
the bare deformation potential is more sensitive to the wave number 
than the screened one;
this is unrelated to the orbital character.
However, it is not a problem when we discuss qualitatively.}.
$Z_{{\rm Ni} 3}$ is larger than other $Z$s when we use the bare vertex
whereas it is smaller than the others when we use the screened vertex.
This result shows that the screening effect on the interaction between the Ni 3{\it d} orbital and phonons
is particularly large;
this strong screening makes the interaction especially weak.
%
There is also some supporting experimental evidence of our scenario.
First, YPd$_2$B$_2$C has the transition temperature higher 
than that of YNi$_2$B$_2$C \cite{ISI:A1994MQ78000051}.
Second, the anomalous behavior of the specific heat of YNi$_2$B$_2$C is reduced
when some Ni atoms are replaced with Pt atoms in the specific-heat measurement
\cite{doi:10.1143/JPSJ.68.1078,Nohara20002177,PhysRevLett.86.1327}. 
According to our scenario, the Pd 4{\it d} orbitals and Pt 5{\it d} orbitals 
are more delocalized than the Ni 3{\it d} orbitals
and this delocalized nature is advantageous to the electron-phonon coupling.
%

We reproduced quantitatively the superconducting $T_c$, the isotope effect constant,
the phonon dispersion excepting the large softening of the TA mode and
reproduced qualitatively the broadened peak structure in the tunnel conductance\cite{PhysRevB.67.014526}
and the $k$ dependence of $\Delta_{n k}$ observed by ARPES\cite{PhysRevB.81.180509}.
However,
the anisotropy of the superconducting gap in our calculation
is too small to reproduce
the ultrasonic attenuation measurement\cite{PhysRevLett.92.147002} and
the magnetic field dependence of the thermal conductivity\cite{PhysRevLett.89.137006}.
We assume one of the origin of this underestimation of the anisotropy to be in
the calculation of the electronic structure in the normal state.
In the previous study of the combination of dHvA experiment\cite{TERASHIMA1995459}
and the band-structure calculation\cite{Yamauchi2004225},
authors shifted upwardly Y 4{\it d} and Ni 3{\it d} levels 
from the LDA levels by 0.11 Ry and 0.05 Ry.
They state these shifts correspond to the self-interaction
and/or the non-local correction to the LDA.
On the other hand, 
reproduction of the Fermi surfaces that agree well with the experiments 
without such an empirical treatment has not been achieved so far.
Thus, the detailed shape of the Fermi surfaces has not been settled.
If we improve on the description of the Fermi surface,
the following may be accomplished.
\begin{itemize}
\item The nesting which corresponds to the TA mode at $q = 0.55 {\rm \Gamma Z_{next}}$ become more significant, 
  yielding stronger softening of the low-energy phonon mode;
  the strength of the nesting is sensitive to the fine structure of the Fermi surface.
\item Regions which consist {\it only} of Ni 3{\it d} orbital appear;
  such regions should couple with phonons very weakly and have quite small gaps.
\end{itemize}

%
%
\section{Summary}
In this study, we performed a first principle investigation
to clarify the origin of the anisotropic superconductivity in YNi$_2$B$_2$C.
We improved the numerical method for the $k$-integration in the
gap equation to treat accurately $k$-dependencies of the electron-phonon interaction
and the gap function.
From calculations with this method,
we found that the anisotropic superconductivity is traced back to
the variation of the rate of the Ni 3{\it d} orbital on the Fermi surface.
As the component of the Ni 3{\it d} orbital increases, 
the electron-phonon coupling of the electronic state becomes weak and 
its superconducting gap function becomes small. Because of this effect, 
the superconducting gap significantly varying over the Fermi surface emerges.
As a possible scenario, we proposed that the localized nature of 
the Ni 3{\it d} orbitals is a key factor for the weakening of the electron-phonon coupling.
We found the relation between the peculiar electron-phonon interaction
and the electronic state in the vicinity of the Fermi surface in this material.

%
%
\begin{acknowledgments}
  We thank Yasutami Takada for his many invaluable advices.
  This work was supported by 
  the Elements Strategy Initiative Center for Magnetic Materials (ESICMM)
  under the outsourcing project of MEXT.
  The numerical calculations were performed using Fujitsu FX10s
  at the Information Technology Center 
  and the Institute for Solid State Physics, The University of Tokyo.
\end{acknowledgments}
\appendix
\section{Frequency integration in Eq. (\ref{fml_kelkernel})}\label{app_omegaint}

In Eq. (\ref{fml_kelkernel}), we perform an integration from 0 to the infinity
as follows:
First we employ a new variable $x$, where
\begin{align}
  \omega = (|\xi_{n k}|+|\xi_{n' k'}|) \frac{1+x}{1-x},
\end{align}
and we obtain 
\begin{align}
  K^{\rm ee}[V_{n' k' n k}](\xi_{n k}, \xi_{n' k'}) =
  \frac{2}{\pi} \int_{-1}^{1} d x \frac{1}{1 + x^2} 
  V_{n k n' k'} \left((|\xi_{n k}|+|\xi_{n' k'}|) \frac{1+x}{1-x}\right).
\end{align}
We use the Gauss-Legendre quadrature for this integration.
To obtain the Coulomb interaction $V_{n k n' k'}(\omega)$ 
at an arbitrary frequency $\omega$,
we employ the Chebyshev interpolation\cite{Press:2007:NRE:1403886}.

\section{Four dimensional numerical integration scheme for DOS}\label{app_pentachron}

We calculate the integration weight $w_i$ in Eq. (\ref{fml_pentachron}) 
as follows,
where $f_{i j}\equiv(\varepsilon - \varepsilon_j)/(\varepsilon_i-\varepsilon_j)$,
$\varepsilon_i$ is $\varepsilon_{k}(\xi)$
at the each corner of a pentachoron.
\begin{enumerate}
\item For $\varepsilon_1$ $<$ $\varepsilon$ $<$ $\varepsilon_2$ $<$ 
  $\varepsilon_3$ $<$ $\varepsilon_4$ $<$ $\varepsilon_5$, we obtain
  \begin{align}
    C &= \frac{f_{2 1} f_{3 1} f_{4 1}}{\varepsilon_5 - \varepsilon_1} 
    \\
    w_1 &= C (f_{1 2} + f_{1 3} + f_{1 4} + f_{1 5}),
    \nonumber \\
    w_2 &= C f_{2 1},
    \hspace{0.5cm}
    w_3 = C f_{3 1},
    \nonumber \\
    w_4 &= C f_{4 1},
    \hspace{0.5cm}
    w_5 = C f_{5 1}.
  \end{align}
\item For $\varepsilon_1$ $<$ $\varepsilon_2$ $<$ $\varepsilon$ $<$ 
  $\varepsilon_3$ $<$ $\varepsilon_4$ $<$ $\varepsilon_5$, we obtain
  \begin{align}
    C_1 &= \frac{f_{3 1} f_{4 1} f_{2 3}}{\varepsilon_5 - \varepsilon_1}, 
    \hspace{0.5cm}
    C_2 = \frac{f_{4 1} f_{3 2} f_{2 4}}{\varepsilon_5 - \varepsilon_1},
    \hspace{0.5cm}
    C_3 = \frac{f_{3 2} f_{4 2} f_{2 5}}{\varepsilon_5 - \varepsilon_1}
    \\
    w_1 &= C_1 (f_{1 3} + f_{1 4} + f_{1 5})
    + C_2 (f_{1 4} + f_{1 5})
    + C_3 f_{1 5},
    \nonumber \\
    w_2 &= C_1 f_{2 3}
    + C_2 (f_{2 3} + f_{2 4}) 
    + C_3 (f_{2 3} + f_{2 4} + f_{2 5}),
    \nonumber \\
    w_3 &= C_1 (f_{3 1} + f_{3 2})
    + C_2 f_{3 2} 
    + C_3 f_{3 2},
    \nonumber \\
    w_4 &= C_1 f_{4 1} 
    + C_2 (f_{4 1} + f_{4 2})
    + C_3 f_{4 2},
    \nonumber \\
    w_5 &= C_1 f_{5 1}
    + C_2 f_{5 1} 
    + C_3 (f_{5 1} + f_{5 2}).
  \end{align}
\item For $\varepsilon_1$ $<$ $\varepsilon_2$ $<$ $\varepsilon_3$ $<$ 
  $\varepsilon$ $<$ $\varepsilon_4$ $<$ $\varepsilon_5$, we obtain
  \begin{align}
    C_1 &= \frac{f_{3 5} f_{2 5} f_{4 3}}{\varepsilon_5 - \varepsilon_1},
    \hspace{0.5cm}
    C_2 = \frac{f_{2 5} f_{3 4} f_{4 2}}{\varepsilon_5 - \varepsilon_1},
    \hspace{0.5cm}
    C_3 = \frac{f_{3 4} f_{2 4} f_{4 1}}{\varepsilon_5 - \varepsilon_1}
    \\
    w_1 &=
    C_1 f_{1 5} 
    + C_2 f_{1 5} 
    + C_3 (f_{1 4} + f_{1 5}),
    \nonumber \\
    w_2 &=
    C_1 f_{2 5} 
    + C_2 (f_{2 4} + f_{2 5})
    + C_3 f_{2 4},
    \nonumber \\
    w_3 &=
    C_1 (f_{3 4} + f_{3 5}) 
    + C_2 f_{3 4} 
    + C_3 f_{3 4},
    \nonumber \\
    w_4 &=
    C_1 f_{4 3}
    + C_2 (f_{4 2} + f_{4 3})
    + C_3 (f_{4 1} + f_{4 2} + f_{4 3}),
    \nonumber \\
    w_5 &=
    C_1 (f_{5 3} + f_{5 2} + f_{5 1})
    + C_2 (f_{5 1} + f_{5 2})
    + C_3 f_{5 1}.
  \end{align}
\item For $\varepsilon_1$ $<$ $\varepsilon_2$ $<$ $\varepsilon_3$ $<$ 
  $\varepsilon_4$ $<$ $\varepsilon$ $<$ $\varepsilon_5$ , we obtain
  \begin{align}
    C &= \frac{f_{4 5} f_{3 5} f_{2 5}}{\varepsilon_5 - \varepsilon_1}
    \\
    w_1 &= C f_{1 5},
    \hspace{0.5cm}
    w_2 = C f_{2 5},
    \nonumber \\
    w_3 &= C f_{3 5},
    \hspace{0.5cm}
    w_4 = C f_{4 5},
    \nonumber \\
    w_5 &= C (f_{5 1} + f_{5 2} + f_{5 3} + f_{5 4}).
  \end{align}
\end{enumerate}

%
%
%

%
\end{document}